\documentstyle[12pt,epsfig,amssymb]{article}
\newskip\humongous \humongous=0pt plus 1000pt minus 1000pt

\newif\ifdtup

\def\Re{\mathop{\rm Re}}

\def\abs#1{\left| #1\right|}

\def\beq{\begin{equation}}
\def\eeq{\end{equation}}

\def\beqn{\begin{eqnarray}}
\def\eeqn{\end{eqnarray}}

\relax

\catcode`\@=11

\@addtoreset{equation}{section}
\def\theequation{\thesection.\arabic{equation}}

\def\@normalsize{\@setsize\normalsize{15pt}\xiipt\@xiipt
\abovedisplayskip 14pt plus3pt minus3pt%
\belowdisplayskip \abovedisplayskip
\abovedisplayshortskip \z@ plus3pt%
\belowdisplayshortskip 7pt plus3.5pt minus0pt}

\def\small{\@setsize\small{13.6pt}\xipt\@xipt
\abovedisplayskip 13pt plus3pt minus3pt%
\belowdisplayskip \abovedisplayskip
\abovedisplayshortskip \z@ plus3pt%
\belowdisplayshortskip 7pt plus3.5pt minus0pt
\def\@listi{\parsep 4.5pt plus 2pt minus 1pt
     \itemsep \parsep
     \topsep 9pt plus 3pt minus 3pt}}

\@twosidetrue

\relax

\catcode`@=12

\catcode`\@=11

\def\section{\@startsection{section}{1}{\z@}{3.5ex plus 1ex minus
   .2ex}{2.3ex plus .2ex}{\large\bf}}
\def\subsection{\@startsection{subsection}{2}{\z@}{3.25ex plus 1ex minus
   .2ex}{1.5ex plus .2ex}{\bf}}

\def\thesection{\arabic{section}}
\def\thesubsection{\arabic{section}.\arabic{subsection}}

\def\appendix{\setcounter{section}{0}
 \def\thesection{Appendix \Alph{section}}
 \def\thesubsection{\Alph{section}.\arabic{subsection}}
 \def\theequation{\Alph{section}.\arabic{equation}}
 \def\section{\@startsection{section}{1}{\z@}{3.5ex plus 1ex minus
   .2ex}{2.3ex plus .2ex}{\large\bf}}
 \def\subsection{\@startsection{subsection}{2}{\z@}{3.25ex plus 1ex minus
   .2ex}{1.5ex plus .2ex}{\bf}}}
\def\marksection{\def\@currentlabel{\Alph{section}}}

\def\ps@headings{\def\@oddfoot{}\def\@evenfoot{}
\def\@oddhead{\hbox{}\hfill
 \makebox[.5\textwidth]{\raggedright\ignorespaces --\thepage{}--
 \hfill {}}}  
\def\@evenhead{\@oddhead}
\def\subsectionmark##1{\markboth{##1}{}}
}

\ps@headings

\catcode`\@=12

\def\figcap{\section*{Figure Captions\markboth
 {FIGURECAPTIONS}{FIGURECAPTIONS}}\list
 {Fig. \arabic{enumi}:\hfill}{\settowidth\labelwidth{Fig. 999:}
 \leftmargin\labelwidth
 \advance\leftmargin\labelsep\usecounter{enumi}}}
 \relax
\def\tablecap{\section*{Table Captions\markboth
 {TABLECAPTIONS}{TABLECAPTIONS}}\list
 {Table \arabic{enumi}:\hfill}{\settowidth\labelwidth{Table 999:}
 \leftmargin\labelwidth
 \advance\leftmargin\labelsep\usecounter{enumi}}}
 \relax
\def\reflist{\section*{References\markboth
 {REFLIST}{REFLIST}}\list
 {[\arabic{enumi}]\hfill}{\settowidth\labelwidth{[999]}
 \leftmargin\labelwidth
 \advance\leftmargin\labelsep\usecounter{enumi}}}
 \relax

\catcode`\@=11

\relax

\catcode `@ 11
\def\biblabel#1{\if@filesw\immediate
\write\@auxout{\string\bibcite{#1}{\the\value{\@listctr }}}\fi}
\catcode `@ 12
\relax
\def\pl#1#2#3{{\it Phys. Lett. }{\bf #1}(19#2)#3}

\def\prep#1#2#3{{\it Phys. Rep. }{\bf #1}(19#2)#3}

\def\np#1#2#3{{\it Nucl. Phys. }{\bf #1}(19#2)#3}

\relax

\newcommand{\ccaption}[2]{
  \begin{center}
    \parbox{0.85\textwidth}{
      \caption[#1]{\small\it {#2}}}
  \end{center}    }
\def    \be             {\begin{equation}}
\def    \ee             {\end{equation}}
\def    \ba             {\begin{eqnarray}}
\def    \ea             {\end{eqnarray}}

\def    \=              {\;=\;}
\def    \frac           #1#2{{#1 \over #2}}

\def\b0{b_0}

\def \qt   {\mbox{$q_{\rm \scriptscriptstyle T}$}}

\def \mt   {\ifmmode m_{\rm t} \else $m_{\rm t}$ \fi}

\def \to   {\mbox{$\rightarrow$}}
\newcommand     \MSB            {\ifmmode {\overline{\rm MS}} \else
                                 $\overline{\rm MS}$  \fi}
\newcommand\hepph[1]{{\tt hep-ph/#1}}

\newcommand\sss{\scriptscriptstyle \rm}
\newcommand\sssrm{\sss \rm}

\newcommand\asq{\alpha^2_{\sss S}}
\newcommand\qtq{q^2_{\sssrm T}}
\newcommand\geul{\gamma_{\sss E}}

\newcommand\Jo{J_1}

\newcommand\sud{{\cal S}}
\newcommand\bsud{{\mathbb S}}
\newcommand\sudNLL{\sud_{\sssrm NLL}}
\newcommand\sudLL{\sud_{\sssrm LL}}

\newcommand\tilsud{\tilde{\cal S}}

\newcommand\bh{\hat{b}}
\newcommand\lb{l_{\hat{b}}}
\newcommand\lbq{l_{\hat{b}}^2}
\newcommand\co{c_2}
\newcommand\coq{c_2^2}
\newcommand\ct{c_1}
\newcommand\ctq{c_1^2}
\newcommand\bz{b_0}

\newcommand\LambdaQCD{\Lambda_{\rm \scriptscriptstyle QCD}}
\newcommand\as{\alpha_s}

\newcommand\CF{C_{\sss F}}
\newcommand\xa{x_{\rm \scriptscriptstyle A}}
\newcommand\xb{x_{\rm \scriptscriptstyle B}}
\newcommand\fxa{f_{q/{\rm \scriptscriptstyle A}}}
\newcommand\fxb{f_{\bar{q}/{\rm \scriptscriptstyle B}}}

\begin{document}
\begin{titlepage}
\nopagebreak

\vskip -2cm

{\flushright{
        \begin{minipage}{4cm}
        CERN-TH/98-278  \hfill \\
        ETH-TH/98-24 \hfill \\
        GEF-TH-9/1998 \hfill \\
        IFUM 630/FT \hfill \\
        hep-ph/9809367\hfill \\
        \end{minipage}        }

}
\vfill
\begin{center}
{\large\sc \baselineskip 0.9cm
  Problems in the resummation of soft-gluon effects in the
  transverse-momentum distributions of massive vector bosons in hadronic
  collisions
}
\vskip .5cm
{\bf Stefano Frixione\footnote{Work supported by the Swiss National
  Foundation.}$^,$\footnote{Address after October 1$^{st}$, 1998: 
TH Division, CERN, Geneva, Switzerland.}}
\\
\vskip 0.1cm
{Theoretical Physics, ETH, Zurich, Switzerland} \\
\vskip .5cm
{\bf Paolo Nason\footnote{On leave of absence from INFN, Milan, Italy.}}
\\
\vskip 0.1cm
{TH Division, CERN, Geneva, Switzerland} \\
\vskip .5cm
{\bf Giovanni Ridolfi}
\\
\vskip 0.1cm
{INFN, Sezione di Genova, Genoa, Italy} \\
\end{center}
\nopagebreak
\vfill
\begin{abstract}
  We consider the resummation of soft-gluon emission in the
  transverse-mo\-men\-tum distribution of vector mesons in hadronic
  collisions. We find that the resummed expression in the
  impact-parameter formulation has an expansion in $\as$ with
  factorially growing terms with oscillating signs.  These diverging
  terms arise from the small impact-parameter region of integration,
  and are of a subleading nature.  We also obtain a closed expression
  for the next-to-leading logarithm resummation in $\qt$-space, and we
  study its analytic structure.  We find in this case that, although
  no factorially growing terms are present, there are geometrical
  singularities that severely restrict the range of applicability of
  the resummation formula.

\end{abstract}
\vfill
CERN-TH/98-278 \hfill \\
September 1998\hfill
\end{titlepage}
\newcommand\mvec{m}
\section{Introduction}

In this paper we discuss soft-gluon resummation effects in the
transverse momentum distributions of heavy systems produced in
hadronic collisions. We will mainly focus on the Drell--Yan process,
but several aspects of the problems we are considering here
are of a general nature, and apply also to
other cases, such as the production of Higgs bosons
or of heavy-quark pairs.

To simplify the notation, we consider the
production of a virtual photon of off-shellness $Q$
(this case can be trivially extended to $W$ or $Z$ production).
We consider the kinematic region where its transverse
momentum $\qt$ is much less than $Q$, but much larger than
$\LambdaQCD$. This regime, characterized by two distinct scales, has been
extensively studied in the past
\cite{DDTpaper,ParisiPetronzio}.  Large logarithms of the ratio
$Q/\qt$ (the so-called Sudakov logarithms) arise to all orders of the
perturbative expansion, with up to two powers of the logarithm for
each power of $\as$.  All-order resummation of the perturbative
expansion is needed in order to make sensible predictions in this
regime. In the small-$\qt$ limit, the resummed cross section has 
the structure suggested by Dokshitzer, Dyakonov and Troyan (DDT)
in ref.~\cite{DDTpaper}:
\beqn
\frac{d\sigma}{dQ^2 d\qtq} &=& \frac{\sigma_0}{Q^2}
\sum_q e_q^2 \frac{d}{d \qtq}  \int d \xa \,d\xb\,\delta(\xa\xb-\tau)\,
\nonumber \\  
&&\times \left\{ \fxa(\xa,\qt)\,\fxb(\xb,\qt)
  \exp\tilsud(Q,\qt)+
  \left(q \leftrightarrow \bar{q} \right)\right\}\,,
\label{DDT}
\eeqn
where $\sigma_0=4\pi\alpha_{em}^2/(9s)$ is the lowest-order cross section 
for the process \mbox{$q\bar{q}\,\to\,\mu^+\mu^-$}, $\tau=Q^2/s$, $s$ is the 
hadronic centre-of-mass energy squared, and
\beq
\tilsud(Q,\qt) = -\int_{\qtq}^{Q^2} \frac{d\mu^2}{\mu^2}
\left[ \tilde{A}(\as(\mu^2)) \log\frac{Q^2}{\mu^2} + \tilde{B}(\as(\mu^2))
\right]
\label{Calessetilde}
\eeq
with
\beq
\tilde{A}(\as)=\tilde{A}_1\as
                 +\tilde{A}_2\as^2+\ldots 
\,,\quad 
\tilde{B}(\as)=\tilde{B}_1\as+\ldots\,.
\eeq
In eq.~(\ref{DDT}), the leading logarithmic (LL) level of accuracy,
which accounts for all terms of order
$\as^n l^{n+1}$ (where $l=\log(Q^2/\qtq)$)
in eq.~(\ref{Calessetilde}),
is achieved by including only the $\tilde{A}_1$ term of the
expansion of $\tilde{A}(\as)$. The next-to-leading logarithmic (NLL)
level, which amounts
to the inclusion of all terms of order
$\as^n l^{n}$, is achieved if one includes also the $\tilde{A}_2$ and
the $\tilde{B}_1$ terms. Using $\qt$ instead of $Q$ in the parton densities
also corresponds to a NLL correction.\footnote{In order to go beyond
the NLL terms, besides the inclusion of higher-order terms in the
expansion of $\tilde{A}(\as)$ and $\tilde{B}(\as)$, it is also
needed to perform a suitable change of scheme on the parton densities.}

We observe that no simple classification of the Sudakov logarithms
can be given by looking at the expansion of the cross section
(\ref{DDT}) in powers of $\as$ and of $l$.
This is because the exponent $\tilsud$ has the formal expansion
in $\as$
\beq
\label{stildeexp}
\tilsud(Q,\qt) = \sum_{n=1}^\infty d_{n,n+1}
   \as^n(Q^2)\,l^{n+1}
 + \sum_{n=1}^\infty d_{n,n} \as^n(Q^2)\,l^n + \ldots\;.
\eeq
When only LL terms are included, the neglected
NLL corrections correspond to a factor, which is a power series
in terms of $\as l$, with the $0^{\rm th}$ order term equal to 1.
Thus, as long as $\as l <1$, the expansion is valid.
On the other hand, by expanding  the exponential of the first
LL term, which is of order $\as l^2$,
we can get arbitrarily large powers of $\as l^2$,
and NLL corrections would appear to be of the order of
$\as^n l^{2n-1}$. In this case, the expansion would seem to be valid
only when $\as l^2 <1$, a much more restricted range.

As we will show later, eq.~(\ref{DDT}) has never appeared
in the literature in a complete form at the NLL level.
Indeed, it was shown in ref.~\cite{ParisiPetronzio} that the resummation
of soft-gluon effects is most naturally performed using the impact-parameter
formalism. A general formula for the Drell--Yan cross section
in the impact-parameter space has been obtained in ref.~\cite{CSSpaper}:
\beqn
&& \frac{d\sigma}{dQ^2 d\qtq} = \frac{\sigma_0}{4\pi Q^2}
\sum_q e_q^2  \int d \xa \,d\xb\,\delta(\xa\xb-\tau)\,
\int d^2 b\, e^{i\vec{q}_{\sssrm T}\cdot \vec{b}}
\nonumber \\ && \phantom{ino}
\times\left\{\fxa(\xa,\ct/b)\,\fxb(\xb,\ct/b)
\exp\sud(Q,b)+
\left(q \leftrightarrow \bar{q} \right)\right\} \,,
\phantom{spingi}
\label{CSS}
\eeqn
where
\beqn
&& \sud(Q,b) = -\int_{\ctq/b^2}^{\coq Q^2}
\frac{d\mu^2}{\mu^2} \left[ A(\as(\mu^2)) \log\frac{\coq Q^2}{\mu^2} +
B(\as(\mu^2)) \right]
\label{suddef} \phantom{spingi}
\\
&& A(\as)=A_1\as+A_2\as^2+\ldots 
\,,\quad 
B(\as)=B_1\as+\ldots\,,
\label{ExpAB}
\eeqn
and $\ct,\co$ are arbitrary constants of order one.\footnote{In
ref.~\cite{CSSpaper}, eq.~(\ref{CSS}) is claimed to be valid also for
$\qt<\LambdaQCD$. This regime is not under investigation here.}
The coefficients $A_i$ and $B_i$ also carry
a dependence upon these arbitrary constants (see \ref{ABcoeff}).
There are well-known problems that arise when applying
eq.~(\ref{CSS}) to obtain phenomenological results. First of all, one
can immediately recognize that, for any value of $\qt$,
eq.~(\ref{CSS}), as it stands, is undefined, because the $b$
integration forces the scale at which $\as$
and the parton densities are evaluated, to approach the
non-perturbative region. This problem is usually avoided by
replacing $b$ in $\sud(Q,b)$ and in the parton densities
with~\cite{CollinsSoperEEC}
\newcommand\blim{b_{\rm lim}}
\beq
\label{blim}
b^\star=\frac{b}{\sqrt{1+(b/\blim)^2}}\,,
\eeq
which approaches $b$ for $b\ll\blim$, and never exceeds $\blim$ for large
values of $b$. Choosing $\blim$ of the order of 1~GeV$^{-1}$, one
thus avoids the non-perturbative region.  This procedure introduces
corrections of the order of powers of $1/\blim$ to eq.~(\ref{CSS}).
Apart from the introduction of $b^\star$, eq.~(\ref{CSS}) is usually
also supplemented with a non-perturbative correction
factor\footnote{In fact, it is argued in ref.~\cite{CSSpaper}
that the non-perturbative region contributes a factor of the form
$\exp[-g_1(b)\log(Q^2)-g_2(b)]$.} of the form $\exp(-gb^2)$. We stress
that this factor is actually not needed in order for the integral
in eq.~(\ref{CSS}) to converge (a proof is given in \ref{nonpert}).

In the present work, we will obtain an explicit expression for eq.~(\ref{DDT}),
valid at the NLL level, and we will address several theoretical problems
in the formulation of eqs.~(\ref{DDT}) and (\ref{CSS}).

\section{$\qt$-space formulation of the resummed cross section}
In this section, we will obtain a $\qt$-space expression for the cross
section given in eq.~(\ref{CSS}), which is valid in the NLL approximation.
We will analitycally perform
the integration over $b$, which appears in eq.~(\ref{CSS}).
We will retain only the terms required for maintaining a NLL accuracy in
the formal power expansion of the cross section. In other words,
we will obtain an expansion for $\tilsud(Q,\qt)$ that corresponds to
retaining the first two sums of eq.~(\ref{stildeexp}), and nothing more.
We will first consider a simplified model, which
corresponds roughly to the assumption of scale-independent
parton densities and coupling constant. This model
can be worked out quite easily, and it
displays some features and problems that persist in the realistic case.
\subsection{A simple model}
\label{sec:model}
We thus assume that the
parton densities and $\as$ do not depend on the energy scale. In this
approximation, eq.~(\ref{CSS}) factorizes in a simple form:
\beq
\frac{d\sigma}{dQ^2 d\qtq} = \frac{\sigma_0}{Q^4}
F(Q,\qt)\,
\sum_q e_q^2  \int d \xa \,d\xb\,\fxa(\xa)\,\fxb(\xb)\,\delta(\xa\xb-\tau),
\label{xsecmodel}
\eeq
where, following ref.~\cite{EllisVeseli}, we have defined the form factor
\beq
F(Q,\qt)=\frac{Q^2}{4\pi}
\int d^2 b\, e^{i\vec{q}_{\sssrm T}\cdot \vec{b}}
\exp\sud(Q,b).
\label{ffmodel}
\eeq
After performing the angular integration in eq.~(\ref{ffmodel}) and changing
the integration variable from $b$ to $\bh=b\qt$ we get
\beq
F(Q,\qt)=\frac{Q^2}{2\qtq}\int_0^{\infty} d\bh\,\bh\, J_0(\bh)\,
\exp\sud(Q,\bh/\qt),
\label{ffdef}
\eeq
where $J_0$ is the $0^{\rm th}$-order Bessel function of the first kind.
Equation~(\ref{ffdef}) can be integrated by parts, exploiting the identity
\beq
x\, J_0(x)=\frac{d}{dx}\,\left(x\, J_1(x)\right).
\label{Besselid}
\eeq
We obtain
\beq
F(Q,\qt)=Q^2\,\frac{d}{d\qtq}\int_0^\infty\,d\bh\,\Jo(\bh)\,
\exp\sud(Q,\bh/\qt)\,.
\label{ffB0}
\eeq
In this way, we have an expression in the DDT form,
analogous to eq.~(\ref{DDT}).

We assume that the Sudakov form factor at the LL level takes the form
\newcommand\aeff{a}
\beq
\sud(Q,\bh/\qt)=-\aeff\,L^2\,,
\label{Sudex0}
\eeq
where
\begin{equation}
  L=\log\frac{Q^2 \bh^2}{\qtq}
\end{equation}
and
\begin{equation}
  \aeff=\frac{A_1}{2}\as\;,
\end{equation}
which is what is obtained if the running of $\as$ is neglected
in eq.~(\ref{suddef}).
In this context we take the arbitrary constants $\ct,\co$ to be equal to~1;
different choices amount to a rescaling of the coupling constant.
Defining
\beq
l=\log\frac{Q^2}{\qtq},\quad \lb=\log\bh^2\,,
\eeq
we obtain
\beq\label{Fmodel}
F(Q,\qt)=Q^2\,\frac{d}{d\qtq}\left\{\exp(-\aeff l^2)
\int_0^\infty d\bh\,J_1(\bh)\exp\left(
-2\aeff l \lb -\aeff \lbq\right)\right\}\;.
\eeq
The exponential under the integral sign can now be expanded
and integrated term by term. In fact, $J_1(\bh)$ has an oscillating behaviour
for large $\bh$, which gives rise to finite integrals when multiplied
by any power of $\lb$. Furthermore, all terms of the expansion
will have at most as many powers of $l$ as powers of $\aeff$.
The NLL level of accuracy is reached by neglecting the $\aeff \lbq$ term
in the exponent, which is formally subleading. We thus get, at the NLL
level,
\beqn
F(Q,\qt)&=&Q^2\,\frac{d}{d\qtq}\left[\exp(-\aeff l^2)
\int_0^\infty\,d\bh\,\Jo(\bh)\,\bh^{-4\aeff l}\right]
\nonumber \\
&=& Q^2\,\frac{d}{d\qtq}\,\left[\exp(-\aeff l^2)\,
2^{-4\aeff l}\;\frac{\Gamma(1-2\aeff l)}{\Gamma(1+2\aeff l)}\right]\;,
\label{ffexNLL}
\eeqn
where the last integral was performed
using the result
\beq
\int_0^\infty\,dx\,x^h\,J_1(x)=2^h\,
\frac{\Gamma(1+h/2)}{\Gamma(1-h/2)},\;\;\;\;-2<h<\frac{1}{2}\,.
\label{int1}
\eeq
The form factor in eq.~(\ref{ffexNLL}) is singular for $\aeff l\, \to\, 1/2$.
Considered as a function of $\aeff$ at a given value of $l$, it is
analytic in a circle of radius $1/(2l)$ around $\aeff=0$.  Thus, when
expanded in powers of $\aeff$, the coefficients of the expansion grow
at most geometrically.

On the other hand, the original expression eq.~(\ref{Fmodel}) is well defined
for any value of $\aeff>0$ and it is divergent for any
negative value of $\aeff$. Thus it cannot be analytic in the
neighbourhood
of the origin. This observation anticipates our conclusions:
the original form factor has a divergent power series expansion in $\aeff$,
but, if we retain only the LL and NLL terms,
we are left with a convergent expansion. Thus, the divergent behaviour
must be ascribed to the subleading terms in the original expression.

We now analyse more closely the origin of the non-analyticity
of eq.~(\ref{Fmodel}), now interpreted as defining
a formal expansion in powers of $\aeff$ (it is easy to see that
the coefficients of this expansion are finite order by order).
First of all, we separate the large-$\bh$
and small-$\bh$ contributions to the integral.
No singularities may arise from the large-$\bh$ region. This can be shown as
follows.
We rewrite the large-$\bh$ part of the integral in eq.~(\ref{Fmodel}) as
\begin{equation} \label{largebh1}
\int_{\bh_0}^\infty d\bh\,J_1(\bh)\exp\left(
-2\aeff l \lb -\aeff \lbq\right)\;=
-\frac{2}{\pi} \Re\left[ \int_{\bh_0}^\infty d\bh
 K_1(-i\bh)\exp\left(
-2\aeff l \lb -\aeff \lbq\right) \right]\;,
\end{equation}
where $\bh_0$ is an arbitrary cut-off.
We can now deform the integration contour counterclockwise, letting
$\bh\,\to\, i\bh$.
Because of the asymptotic behaviour
\begin{equation} \label{k1asympt}
  K_1(\bh) \approx \sqrt{\frac{\pi}{2 \bh}}\, e^{-\bh}\,,
\end{equation}
the integral is strongly convergent, with all its derivatives with respect
to $\aeff$, regardless of the sign of $\aeff$. It thus defines an analytic
function with infinite radius of convergence.

We now study the small-$\bh$ contribution to the integral.
Using the small-$\bh$ behaviour $J_1(\bh)\approx \bh/2$, we get
\begin{equation}
\int_0^{\bh_0} d\bh\,\bh \exp\left(
-2\aeff l \lb -\aeff \lbq\right)
= \frac{\bh_0^2}{2} e^{\aeff l^2} \sum_{n=0}^\infty
\aeff^n  \sum_{i=0}^{2n} (-1)^{n-i} \frac{(2n)!}{n! i!} (l+l_{\bh_0})^i\;.
\end{equation}
We see that subleading terms in this
expansion have factorially growing coefficients. For example, the terms
corresponding to $i=0$ grow as $(2n)!/n!$. The factorial growth
is accompanied by
sign oscillations. This is why, when $\aeff>0$, this asymptotic expansion
corresponds to a finite expression.

In summary, we started with an expression that had an essential singularity
at $\aeff = 0$, arising from the small-$\bh$
region of integration, but well defined for all positive values
of $\aeff$. By dropping all terms beyond the leading and
next-to-leading, we obtained an expression that is analytic
around the origin, but with poles at positive, finite values 
of $\aeff$. We should now give our judgement, and decide which one
of the two expressions is the most reliable.
It is clear from the start that the factorially growing terms in our
initial expression have no justification. It has been pointed out
\cite{ERV} that, in fact, the expression for the Sudakov
form factor should carry a theta function
\begin{equation}\label{thetasuda}
  \sud(Q,b) \,\to\, \theta(b-1/Q) \sud(Q,b)\,,
\end{equation}
since no large logarithms of $b$ arise from the small-$b$ region
in the exact calculation of the first-order correction to the Drell--Yan
cross section.
This theta function would suppress the integrand in the region
$0<\bh<\qt/Q$. It would therefore amount to a power-suppressed
correction that removes the small-$\bh$ singularities.
In fact
\begin{equation}
  \int_0^\infty e^{-aL^2\, \theta(\bh-\qt/Q)} \Jo(\bh)\, d\bh
= \int_0^\infty e^{-aL^2} \Jo(\bh)\, d\bh
- \int_0^{\qt/Q} \left[ e^{-aL^2} -1 \right]\,  \Jo(\bh)\, d\bh\;.
\end{equation}
If we expand the exponential in the first term,
and integrate term by term, we find a power expansion in $a$
whose coefficients are polynomials in $\log \qt/Q$.
For the second term, the same procedure yields an
expansion whose coefficients are suppressed by at least
two powers of $\qt/Q$, as can be seen by expanding the function
$\Jo$ for small values of its argument. The factorial growth
present in the second term compensates the one present in the
first one. The fact remains, however, that if we drop all
power-suppressed effects from our expression, the factorial growth remains.

It should also be pointed out that the replacement given in
eq.~(\ref{thetasuda}) is actually justified only for the
first-order expansion of the Sudakov exponential. In fact, factorization
of soft-gluon emission has a double origin. It comes from the dynamical
factorization of the soft-gluon emission amplitudes, and from the
kinematical factorization of the phase space for the emitted gluons.
In the computation of the transverse momentum of DY pairs, the phase
space factorizes naturally in the impact-parameter (i.e. in $b$)
representation. However, neither the phase space nor the emission
amplitudes factorize for the emission of hard gluons.
To be more specific, while it seems reasonable to assume
that gluon emission is cut off for gluons with transverse momentum
larger than $Q$ (i.e. for $b<1/Q$), for two-gluon emission
this constraint should become stronger, something of the form
$\abs{\qt_1}+\abs{\qt_2} <Q$, and so on.
It is therefore quite possible that the factorial growth coming
from the small-$\bh$ region is only due to an improper treatment
of the multigluon emission for large transverse momenta.
Since the factorial growth is accompanied by sign alternation,
it leads to a suppression of the small-$\bh$ region, thus giving
a more stable result, which is however not fully justified.

It is interesting to compare what we have found here with
what was found in the case of resummation of threshold-enhanced
Sudakov effects \cite{cmnt}. In that case, it had been found that
factorial growth of the perturbative expansion was present
in the $x$-space formulation of the resummation. Its
origin is in the small-momentum region, and is caused by the fact
that momentum conservation is violated in the $x$-space formalism.
If one instead uses the Mellin transform (i.e. $N$-space) formalism,
which correctly implements momentum conservation,
no factorial growth is present. Kinematic factorization
of soft-gluon emission is present in this case in the $N$-space
formulation. It then turns out that no factorial growth is present
when the resummation is performed in the most natural formalism.

In the case we are considering here, instead, the factorial growth
arises in the most natural formulation, which is the $b$-space
formulation. However, it does arise from the large-momentum region,
which is unreliably treated by the resummation formalism.

It is generally believed that factorial growth in the perturbative expansion
may arise from ultraviolet or infrared renormalons.
In the present case, as in the case of ref.~\cite{cmnt},
the factorial growth does not arise from renormalons, since
it is present also at fixed $\as$. In both cases, the factorial growth
appears to be an artefact of the formulation. This is reassuring,
since in most studies of power-suppressed effects it is assumed
that no other sources of factorial growth exist besides renormalons and
instantons.

\subsection{The realistic case}
The realistic case is more complicated, because the parton densities
carry a dependence on the scale $1/b$, which prevents from factorizing
the cross section as in eq.~(\ref{xsecmodel}). We therefore proceed
as in ref.~\cite{EllisVeseli} and take the Mellin transform with respect
to $\tau$ (at fixed $Q^2$) of eq.~(\ref{CSS}):
\beq
\sigma_N=
\int_0^1 d\tau\,\tau^{N-1}\,\frac{Q^2}{\sigma_0}\frac{d\sigma}{dQ^2 d\qtq}.
\eeq
We thus find
\beq
\sigma_N=\frac{1}{4\pi}\int d^2 b \,
e^{i\vec{q}_{\sssrm T}\cdot\vec{b}}\,
\,C_N(\ct/b)\;\exp\sud(Q,b)\;,
\label{sigman}
\eeq
where
\beq
C_N(\mu)=\sum_q e_q^2  \int_0^1 d \xa\,\xa^{N-1}\,\fxa(\xa,\mu) 
\int_0^1 d \xb\,\xb^{N-1}\,\fxb(\xb,\mu)\;.
\label{CNdef}
\eeq
We now use the parton-density evolution equation to write\footnote{In order
to simplify the discussion, we only consider the non-singlet case here.}
\beq
C_N(\ct/b)=C_N(\co Q)\exp{\cal G}_N(Q,b)
\label{PDFEvolution}
\eeq
with
\beq
{\cal G}_N(Q,b)=-2\int_{\ctq/b^2}^{\coq Q^2}
\frac{d\mu^2}{\mu^2}\gamma_N(\as(\mu^2))\;.
\label{GNdef}
\eeq
The anomalous dimensions of the parton-density functions have a perturbative
expansion given by
\beq
\gamma_N(\as)=\gamma_{1,N}\as+\gamma_{2,N}\as^2+\ldots\,.
\eeq
We have therefore
\begin{equation}
\sigma_N=\frac{1}{Q^2}C_N(\co Q) F_N(Q,\qt),  
\end{equation}
where
\beq
F_N(Q,\qt)=\frac{Q^2}{4\pi}\int d^2 b \,
e^{i\vec{q}_{\sssrm T}\cdot\vec{b}}\,
\exp\sud(Q,b)\,\exp{\cal G}_N(Q,b).
\label{ffn}
\eeq
The form factor $F_N(Q,\qt)$ can be computed in the same way as $F(Q,\qt)$
in eq.~(\ref{ffmodel}). We thus obtain the analogue of eq.~(\ref{ffB0}):
\beq
F_N(Q,\qt)
=Q^2\,\frac{d}{d\qtq}\int_0^\infty\,d\bh\,\Jo(\bh)\,\exp\sud(Q,\bh/\qt)
\,\exp{\cal G}_N(Q,\bh/\qt).
\label{ffBN}
\eeq
In this case, however, the integrand in eq.~(\ref{ffBN}) is not defined at
large $\bh$.  In fact, for large $\bh$ the lower integration bound in
eqs.~(\ref{suddef}) and (\ref{GNdef}) decreases and eventually, for some value
of $\bh$, $\as(\mu^2)$ in the integrand is computed at the Landau pole.
For the moment, we consider eq.~(\ref{ffBN}) as a representation
of its formal power expansion in $\as(Q^2)$. The Landau pole
contributes to spoil the convergence of this expansion.
We will study its convergence properties in \ref{AppCochise}.

From eqs.~(\ref{suddef}) and (\ref{ExpAB}) we can obtain
an explicit expansion for $\sud(Q,b)$, which turns out to have the form
\beq
\sud(Q,b)=\sum_{i=0}^\infty\,\as^{i-1}\,f_i(\as\,L),
\quad \as=\as(\coq Q^2)\,,
\quad L = \log\frac{\coq Q^2 \bh^2}{\ctq\qtq}\,.
\label{sudL}
\eeq
In the following we will consider the NLL expansion (in terms of $L$), which
corresponds to
\begin{equation}
  \sud(Q,b)\simeq\bsud(\as,L)\equiv\frac{1}{\as} f_0(\as L)+f_1(\as L)\;,
\end{equation}
where
\beqn
&&f_0(y)=\frac{A_1}{b_0^2}\,\left[b_0 y+\log(1-b_0 y)\right]
\label{leading}
\\
&&f_1(y)=\frac{A_1 b_1}{b_0^2}\,
\left[\frac{1}{2}\log^2(1-b_0 y)
+\frac{b_0 y}{1-b_0 y}+\frac{\log(1-b_0 y)}{1-b_0 y}\right]
\nonumber\\
&&\phantom{f_1(b_0 y)=}
-\frac{A_2}{b_0^2}\,\left[\log(1-b_0 y)+\frac{b_0 y}{1-b_0 y}\right]
+\frac{B_1}{b_0}\,\log(1-b_0 y),
\label{nleading}
\eeqn
and $b_0,b_1$ are the first coefficients of the QCD $\beta$ function:
\beq
\mu^2\frac{\partial\as(\mu^2)}{\partial\mu^2}=-b_0\asq(1+b_1\as+\ldots),
\eeq
\beq
b_0=\frac{33-2 n_f}{12\pi};\;\;\;
b_1=\frac{1}{2\pi}\frac{153-19n_f}{33-2n_f}.
\eeq
Observe that $f_0(y)$ is of order $y^2$, and $f_1(y)$ is of order $y$
as $y\,\to\, 0$.  We expect the remaining $f_i(y)$ to have a power
expansion in $y$ around $y=0$.

Equation~(\ref{sudL})
represents a logarithm expansion of the form factor $\sud$ {\it in
terms of} $L$.  However, as previously observed,
the large logarithm of our physical problem
is $l$ rather than $L$, which depends on the integration variable $\bh$.
The leading and next-to-leading logarithm expansions of the 
form factor $\sud$ in terms of $l$ are
\beqn
\sudLL(Q,\bh/\qt)&=&\frac{1}{\as}f_0(\as\,l),
\label{appLL}
\\
\sudNLL(Q,\bh/\qt)&=&\frac{1}{\as}f_0(\as\,l)+f_1(\as\,l)+
\frac{df_0(\as\,l)}{d(\as\,l)}\,\lb\,.
\label{appNLL}
\eeqn
On the other hand, ${\cal G}_N$ is a pure NLL term, since
the scale dependence of parton densities gives rise to corrections,
which have the form of a power series in $\as l$. Furthermore,
from the definition given in eq.~(\ref{GNdef}), we have
\beq
{\cal G}_N(Q,\bh/\qt)=
{\cal G}_N(Q,1/\qt)+\mathrm{NNLL\; terms}.
\eeq

We first consider the LL approximation. In this case the $\bh$ integration
is trivial, because $\exp\sud$ is independent of $\bh$ at the LL level,
as is clear from eq.~(\ref{appLL}). The factor in eq.~(\ref{GNdef}) 
being a pure NLL correction, we therefore find
\beq
F_N(Q,\qt)=Q^2\frac{d}{d\qtq}
\exp\sudLL\,,
\label{ffintLL}
\eeq
which is, up to NLL terms, the result of ref.~\cite{EllisVeseli}.

We now turn to the NLL approximation. Using eq.~(\ref{appNLL}) we have
\beq
\exp\sudNLL(Q,\bh/\qt)=\left(\frac{\co\bh}{\ct}\right)^h
\exp\bsud(\as,l),
\label{esudNLL}
\eeq
where 
\begin{equation}
h=2\,\frac{df_0(\as\,l)}{d(\as\,l)}
=-\frac{2 A_1}{\bz}\frac{\bz\as l}{1-\bz\as l}.
\label{hdef}
\end{equation}
Since at NLL ${\cal G}_N$ is independent of $\bh$, we have 
\beqn
&&F_N(Q,\qt)=Q^2\frac{d}{d\qtq}\,\left[
\exp{\cal G}_N(Q,1/\qt)\;
\exp\bsud(\as,l)\;
\left(\frac{\co}{\ct}\right)^h
\int_0^\infty\,d\bh\,\bh^h\,J_1(\bh)\right]
\nonumber
\\
&&\phantom{aaaaaa}=Q^2\,\frac{d}{d\qtq}\,\left[\exp{\cal G}_N(Q,1/\qt)
\;\exp\bsud(\as,l)\;
\left(\frac{2\co}{\ct}\right)^h
\frac{\Gamma(1+h/2)}{\Gamma(1-h/2)}\right].
\label{IBres}
\eeqn
Equation~(\ref{IBres}) represents our final result for the form factor
$F_N(Q,\qt)$ at NLL. It can be used to obtain an expression similar
to eq.~(\ref{DDT}) for the cross section; in fact, using the procedure
outlined in ref.~\cite{EllisVeseli}, we may replace eq.~(\ref{IBres})
in eq.~(\ref{sigman}) and obtain
\beqn
\sigma_N&=&
\,\frac{d}{d\qtq}\,\left[C_N(\co  Q)\, 
\exp{\cal G}_N(Q,1/\qt)
\;\exp\bsud(\as,l)\;
\left(\frac{2\co}{\ct}\right)^h
\frac{\Gamma(1+h/2)}{\Gamma(1-h/2)}\right]
\nonumber\\
&=&
\,\frac{d}{d\qtq}\,\left[C_N(\ct\qt)\,
\exp\bsud(\as,l)\;
\left(\frac{2\co}{\ct}\right)^h
\frac{\Gamma(1+h/2)}{\Gamma(1-h/2)}\right],
\eeqn
where we have used eq.~(\ref{PDFEvolution}). The $N$ dependence has
been completely
absorbed in the factor $C_N(\ct\qt)$, so that the inverse Mellin
transform can be
performed immediately and the cross section takes the form
\beqn
&& \frac{d\sigma}{dQ^2 d\qtq} = \frac{\sigma_0}{Q^2}
\sum_q e_q^2 \frac{d}{d \qtq}  \int d \xa \,d\xb\,\delta(\xa\xb-\tau)\,
\nonumber \\&&
\times \left\{ \fxa(\xa,\qt)\,\fxb(\xb,\qt)
\exp\bsud(\as,l)\;
\left(\frac{2\co}{\ct}\right)^h
\frac{\Gamma(1+h/2)}{\Gamma(1-h/2)}
+\left(q \leftrightarrow \bar{q} \right)\right\},
\phantom{spingi}
\label{ourDDT}
\eeqn
where $h$ is given in eq.~(\ref{hdef}).
Equation (\ref{ourDDT}) gives an
explicit expression of the resummed transverse-momentum distribution
of DY pairs at NLL accuracy. This result bears some similarities with
that of ref.~\cite{RakowWebber}, but is in fact different. The
difference arises because we insist on keeping leading and
next-to-leading terms according to the classification described in the
Introduction, while in ref.~\cite{RakowWebber} the emphasis is on
finding a good appoximation to the $b$-space result.
A formula analogous to ours, but for the case
of the energy-energy correlation in $e^+e^-$ annihilations,
can instead be found in ref.~\cite{Turnock}.

As in the case of the simple example treated in the previous
subsection, our expression is analytic in $\as$ for small $\as$,
and displays a pole at $h=-2$, which corresponds to
\begin{equation}
  \as = \frac{1}{l(A_1+\bz)}\;.
\end{equation}
This can also be seen as a lower limit on the allowable values
of $\qt$:
\begin{equation}
  \qt > Q\exp\left(-\frac{1}{2\as(A_1+\bz)}\right)
\simeq 
\LambdaQCD \left(\frac{Q}{\LambdaQCD}\right)^{\frac{A_1}{A_1+\bz}}\;.
\label{Bqtcond}
\end{equation}
We thus find the disappointing result that our formula 
breaks down at a value of $\qt$ that is an
increasing function of $Q/\LambdaQCD$.

The full $b$-space expression is instead well defined for all positive values
of $\as$. However, it is not analytic around the origin.
We will now prove that the small-$\bh$ region contributions to the coefficients
of its power expansion are at least as divergent
as in the simple example
examined in the previous subsection. We thus consider the integral
\begin{equation}
  \int_0^1 d\bh\, \bh\, \exp\left\{\frac{1}{\as \bz^2}
   \left[\as \bz L +\log(1-\as \bz L)\right]\right\}
\end{equation}
with $L=\log (Q^2 \bh^2/\qt^2)$,
where, since we are considering the small-$\bh$ region, we have approximated
$J_1(\bh)\approx \bh/2$. We have fixed for simplicity $\bh_0=1$.
We will focus on the most subleading terms, that is to say, those carrying
no powers of $\log (\qt/Q)$ in the expansion. These terms are obtained
by simply taking $\qt=Q$. Expanding the exponent in powers of $\as$,
our integral becomes
\begin{eqnarray}
&&  \int_0^1 d\bh \,\bh\,\exp\left\{\frac{1}{ \bz^2}
   \sum_{k=2}^\infty \frac{(-\as)^{k-1} \bz^k (-\lb)^k}{k} \right\}
\nonumber \\
&& = \int_0^1 d\bh\, \bh\, \sum_{j=0}^\infty \frac{(-\as)^j}{j!} \left[
\frac{1}{2^j}(-\lb)^{2j}
  +r_1  (-\lb)^{2j-1} + \ldots \right]\;,
\end{eqnarray}
where the highest power in $\lb$ comes from the exponentiation
of the $k=2$ term in the
exponent, whose power of $\lb$ is twice the corresponding power of $\as$.
The important thing to notice is that all the remaining coefficients
$r_1$, $r_2$, etc., are positive. Thus each term of the sum is positive,
and the value of each coefficient is bounded from below by the value of
the first term $\int_0^1 d\bh\, \bh\, (-\lb)^{2j} = (2j)!/2$, which shows the
presence of factorial growth.

In order to complete our task, we should show that no factorial
growth arises from the large-$\bh$ region of integration
\begin{equation}
  \int_1^\infty d\bh J_1(\bh) \exp\left\{\frac{1}{\as \bz^2}
   \left[\as \bz L +\log(1-\as \bz L)\right]\right\}\,.
\end{equation}
Unlike the case of the simple model, here we can no longer rely
upon the analyticity in $\as$ of this integral.
In fact, the logarithm in the exponent encounters a Landau pole
at $\as\bz L=1$, and thus analyticity in $\as$ cannot possibly be there.
Thus, the coefficients of its perturbative expansion will grow faster
than any power, but it is still possible to show that they grow
less than factorially, by using the fact that the above integral
can be related to an exponentially dumped one. The demonstration,
similar to the analogous one in ref.~\cite{cmnt}, is given in
\ref{AppCochise}.

\section{Comparison of various approaches}
\begin{figure}
\begin{center}
    \mbox{
      \epsfig{file=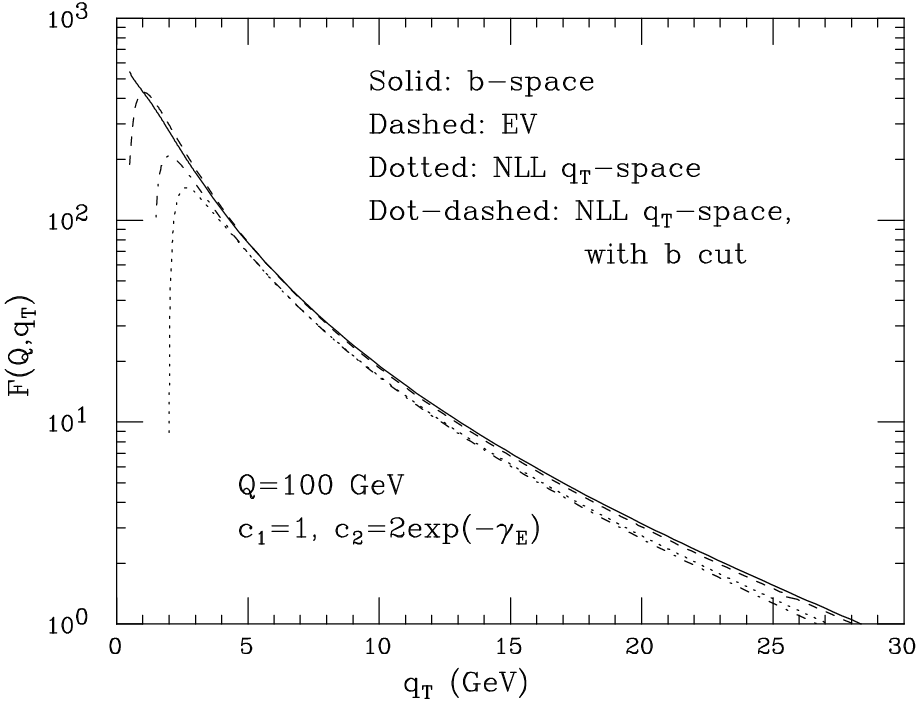,width=0.70\textwidth}
      }
\ccaption{}{\label{b-and-qt}
Comparison of the numerical results of the various approaches 
for the calculation of the form factor.
}
\end{center}
\end{figure}

We will now compare the numerical differences of
the various approaches for the computation of the Sudakov form factor.
For the purpose of comparison, we will consider
eq.~(\ref{ffBN}) without the $N$-dependent term
(which is usually absorbed into a scale
change in the parton densities)
\begin{equation} \label{FormFac}
F(Q,\qt)
=Q^2\,\frac{d}{d\qtq}\int_0^\infty\,d\bh\,\Jo(\bh)\,\exp\sud(Q,\bh/\qt)\,.
\end{equation}
We will consider the following approaches:
\begin{itemize}
\item
  The full $b$-space formula at NLL, 
  that is eq.~(\ref{FormFac}) with $\sud(Q,b)$ given by
  \begin{equation}
    \sud(Q,b)=\bsud(\as,L^\star)=\frac{1}{\as}f_0(\as L^\star)
    +f_1(\as L^\star)\,,
  \end{equation}
  and
  \begin{equation}
    L^\star=\log\frac{\coq  Q^2 {b^\star}^2}{\ctq}\,,\quad\quad
    b^\star=\frac{b}{\sqrt{1+(b/\blim)^2}}\;. 
  \end{equation}
  The functions $f_0$ and $f_1$ are given in eqs.~(\ref{leading}) and
  (\ref{nleading}).
  Here we choose $\blim$ at the position that corresponds to the Landau pole.
\item
  The NLL $\qt$-space formula
  \begin{equation}
    F(Q,\qt)=
    Q^2\,\frac{d}{d\qtq}\,\left[
      \exp\bsud(\as,l)\,
      \left(\frac{2\co}{\ct}\right)^h
      \frac{\Gamma(1+h/2)}{\Gamma(1-h/2)}\right]\;.
    \label{ourF}
  \end{equation}
\item 
  The expression of ref.~\cite{EllisVeseli}. We wish to note at this
  point that the expression of ref.~\cite{EllisVeseli}
  differs from our eq.~(\ref{ourF}) by next-to-leading terms.
  In fact, the expression of ref.~\cite{EllisVeseli} is accurate at the
  leading-logarithmic level, with the further inclusion of subleading
  logarithmic terms down to the order $\as^j l^{2j-2}$ {\it in the
    expansion of the form factor}. This is unlike our approach, in which
  terms down to the order $\as^j l^j$ are kept {\it in the
    Sudakov exponent}.
\end{itemize}
The results of the various approaches are summarized in
fig.~\ref{b-and-qt}.
We see that the $b$-space, the NLL $\qt$-space, and the EV approaches
are quite similar for $\qt\ge 3\;$GeV. Below $3$ GeV, the NLL $\qt$-space
formula begins to display a significant deviation, due to the fact that
it is approaching the singularity $h\to -2$.
As an illustration, we also plot a modification of the NLL $\qt$-space
formula, obtained by letting
\begin{equation}
\label{theta}
  \sudNLL\, \to \,\theta(b-1/Q) \sudNLL=\theta(\bh-\qt/Q) \sudNLL
\end{equation}
before performing the $\bh$ integration (the relevant formulae are
given in \ref{rocut}).
Modified in this way, we see that the NLL result is stabilized down to smaller
values of $\qt$.

The result of ref.~\cite{EllisVeseli} (the dashed curve labelled EV in
the figure) is better behaved at small $\qt$, and much closer to the
full $b$-space formula.  In fact, the EV result differs from the
$b$-space formula by next-to-leading terms. It is paradoxically the
lack of these terms (which would generate the singularity) that
stabilizes the result.

\section{Conclusions}
In summary, we have examined the impact-parameter formula for
the transverse-momentum distribution in Drell--Yan-like processes.
We have found that, when the resummed expression is expanded
in powers of the strong coupling constant, and all terms suppressed
by powers of $\qt/Q$ are neglected, the coefficients of the expansion
exhibit factorial growth with oscillating signs. These diverging terms
arise from the small-impact-parameter region of integration.

The origin of these terms is analogous to what was found in the case of
resummation of threshold-enhanced Sudakov effects \cite{cmnt}. In that
case, however, factorially growing terms arise in the $x$-space formulation of
the resummation, from the small-momentum region of integration. No
factorial growth is present in the $N$-space formulation, which is
most natural in this case, since it preserves momentum conservation.

In the case we are considering here, instead, the factorial growth
arises in the most natural formulation, which is the $b$-space
formulation. It does, however, arise from the large-momentum region,
which is unreliably treated by the resummation formalism.  Presumably,
its origin is due to the fact that energy constraints are not
correctly implemented by the resummation formula for multigluon
emission of sizeable energy.

In both cases, the factorial growth is only present in subleading
(in the logarithmic sense) terms, which are not reliably given
by the resummation formula.

We have obtained an analytic expression for the resummed cross section
in which only the LL and NLL terms in the Sudakov exponent are retained. 
Thus, factorially growing terms are not present in this formula.
However, the formula displays geometric singularities at finite values of
$\as \log Q/\qt $, which severely restrict its range of applicability.
These geometric singularities arise from the small-$b$ region of integration.
In the original $b$-space formula, the small-$b$ region gives rise to
factorially growing terms, and thus to zero radius of convergence.
This factorial growth is accompanied by sign
oscillation in this formulation, and thus by strong dumping of the
small-$b$ region. Thus, the formula is well defined for positive
values of the coupling constant.
Since, however, the factorial growth is an artefact
of the $b$-space formulation, its superior stability (as far as
the small-$b$ region is concerned) may hide some real problems.

We believe that a more realistic treatment of the factorization properties
of mul\-ti\-gluon emission at large transverse momenta could solve
many of the problems discussed so far,
and would thus be a valuable
improvement of the computation of the Drell--Yan transverse-momentum
spectra.

\section*{Acknowledgements}
We thank S.~Catani, M.~Ciafaloni, J.C.~Collins, R.K.~Ellis,
D.~Soper, G.~Sterman and B.R.~Webber for helpful discussions.

\appendix
\section{Coefficients of the Sudakov exponent}
\label{ABcoeff}
In the case of DY pair production, the coefficients
in eq.~(\ref{ExpAB}) are given by \cite{DWS,ExpAB}
\beqn
&&A_1=\frac{\CF}{\pi}
\\
&&A_2=\frac{1}{\pi^2}\left( \frac{67}{9}-\frac{\pi^2}{3}-\frac{10}{27} n_f
   +\frac{8\pi}{3}\,b_0\,\log\frac{\ct e^{\geul}}{2} \right) 
\\
&&B_1=\frac{2\CF}{\pi}\, \log\frac{\ct e^{\geul-3/4}}{2 \co }\,.
\eeqn
The coefficient $B_2$ has also been computed:
\beqn
&&B_2=\frac{1}{\pi^2}
\left[2 \left(\frac{67}{9}-\frac{\pi^2}{3}-\frac{10}{27} n_f\right)
\log\frac{\ct e^{\geul-3/4}}{2 \co }\right.
\nonumber\\
&&\phantom{B_2=}
       +\frac{8\pi}{3}\,b_0 \left(\log^2\frac{\ct e^{\geul}}{2}
        -\log^2 \co  e^{3/4}\right)
\nonumber\\
&&\phantom{B_2=}\left.
-\frac{9}{8}+\frac{7\pi^2}{6}
        +\frac{2\zeta_3}{3}
       +\left(\frac{5}{36}-\frac{2\pi^2}{27}\right) n_f\right]\;.
\eeqn
However, it only contributes at the NNLL level, together with unknown
contributions from the $A_3$ term.

\section{$\qt$-space result with a $b$ cut}
\label{rocut}
Our starting point is eq.~(\ref{FormFac}), with
\begin{equation}
  \sud(Q,\bh/\qt) \,\to\, \theta(\bh-\rho) \sudNLL(Q,\bh/\qt)\,,
\end{equation}
where $\rho=\qt/Q$. We obtain
\begin{eqnarray}
  F(Q,\qt)&=&Q^2\frac{d}{d\qtq}\int_0^\infty d\bh\,\Jo(\bh)\,
         \exp\left[\theta(\bh-\rho)\sudNLL(Q,\bh/\qt)\right]
\nonumber \\
 &=& Q^2\frac{d}{d\qtq}\Bigg\{ \int_0^\rho d\bh\, \Jo(\bh)
      \left[1-\exp\sudNLL(Q,\bh/\qt)\right]
   \nonumber \\ &&
     +\int_0^\infty d\bh\, \Jo(\bh)\exp \sudNLL(Q,\bh/\qt) \Bigg\}\;.
\end{eqnarray}
The second term in the curly bracket corresponds to eq.~(\ref{ourF}).
The first term is easily computed by expanding the Bessel function
\begin{equation}
  \Jo(\bh)=\sum_{k=1}^\infty C_k \bh^k = \frac{\bh}{2} 
  - \frac{\bh^3}{16}+\ldots \;. 
\end{equation}
We obtain
\begin{eqnarray}
  F(Q,\qt)&=&Q^2\frac{d}{d\qtq}\Bigg\{
      \sum_{k=1}^\infty C_k \left[\frac{\rho^{k+1}}{k+1}
       - \exp\bsud(\as,l) 
          \left(\frac{\co}{\ct}\right)^h\frac{\rho^{h+k+1}}{h+k+1}
            \right]\nonumber \\&&
     + \exp\bsud(\as,l)\,
   \left(\frac{2\co}{\ct}\right)^h
    \frac{\Gamma(1+h/2)}{\Gamma(1-h/2)}
  \Bigg\}\;.
\end{eqnarray}
Notice that the term with $k=1$ in the sum has a singularity for $h=-2$,
which cancels exactly the analogous singularity arising in the second term
from $\Gamma(1+h/2)$. Similarly, the
singularities for $h=-4,\,-6\,,\ldots$ are cancelled by the
higher-order terms in the sum.

\section{Growth of the resummed expansion}
\label{AppCochise}
In this Appendix, we show that the large-$\bh$ region of the integral
\begin{equation} \label{Largebint}
  \int_{\bh_0}^\infty d\bh J_1(\bh) I(\as,L)
\end{equation}
with
\begin{equation}
I(\as,L) =  \exp\left\{\frac{1}{\as \bz^2}
   \left[\as \bz L +\log(1-\as \bz L)\right]\right\}
\end{equation}
does not generate factorially growing coefficients in the perturbative
expansion.

In the following, we will use the notation
\begin{equation}
  \sum_{j,k} C_{j,k} (-\as)^j (-L)^k \le \sum_{j,k} G_{j,k} (-\as)^j (-L)^k 
\end{equation}
to mean that
\begin{equation}\label{ourlessthan}
  \abs{C_{j,k}} \le \abs{ G_{j,k}}
\end{equation}
for all values of $j$ and $k$.
The minus sign in front of $\as$ and $L$ above is actually
irrelevant, and it is there only to make the following discussion
more transparent.
We have
\begin{equation}
I(\as,L) = \exp\frac{\as \bz L +\log(1-\as \bz L)}{\as \bz^2}
=\exp \sum_{k=2}^\infty \frac{\bz^{k-1}(-\as)^{k-1} (-L)^k}{\bz k}\,.
\end{equation}
The coefficients of  $(-\as)^{k-1}(-L)^k$ in the power series in the exponent
are all positive, and the exponential function has a power expansion
with positive coefficients. Thus, if we increase the coefficients
of $(-\as)^{k-1}(-L)^k$ in the exponent, we will surely obtain
an expression that is larger (in the sense of eq.~(\ref{ourlessthan}))
than the original one. We thus replace $k\,\to\, k-1$ in the denominators
of the coefficients, and obtain
\begin{equation}
I(\as,L) \le
\exp \sum_{k=2}^\infty \frac{\bz^{k-1}(-\as)^{k-1} (-L)^k}{\bz (k-1)}
 = (1-\as\bz L)^y\,.
\end{equation}
where $y=L/\bz$.
We now continue with the trivial inequalities
\begin{eqnarray}
I(\as,L) & \le & 
 (1-\as\bz L)^y
\nonumber \\
& = &\sum_{k=0}^\infty (- \as \bz L)^k \frac{y(y-1)\ldots (y-k+1)}{k!}
\nonumber \\
& \le & \sum_{k=0}^\infty (\as \bz L)^k \frac{(y+k)^k}{k^k 2^{-k}}
 \le \sum_{k=0}^\infty (2 \as \bz^2)^k (y+1)^{2k}\;,
\end{eqnarray}
where we have used the property
\begin{equation}
  k! > k^k 2^{-k} \;.
\end{equation}
We thus have to examine the asymptotic behaviour of the integral
\begin{equation}
  \int_{\bh_0}^\infty d\bh\,\Jo(\bh) (y+1)^{2k}\,,
\end{equation}
which in turn is that of an integral of the form\footnote{
Here we use the same method as was used in eq.~(\ref{largebh1}),
and also use the fact that only the large-$\bh$ region can lead to
stronger-than-geometric growth.}
\begin{equation}
  \int_{\bh_0}^\infty d\bh \exp(-\bh) \log^{2k} \bh \;.
\end{equation}
Using saddle-point methods, this is easily seen to grow like
$\log^{2k} 2k$, which is stronger than geometric,
but much less than factorial.

\section{A few remarks on the non-perturbative part}
\label{nonpert}
We consider here the $b$-space resummation expression
\begin{eqnarray}
&& \frac{d\sigma}{dQ^2 d\qtq} = \frac{\sigma_0}{Q^2}
\sum_q e_q^2  \int d \xa \,d\xb\,\delta(\xa\xb-\tau)\,
\int d^2 b\, e^{i\vec{q}_{\sssrm T}\cdot \vec{b}}
\nonumber \\ && \phantom{ino}
\times\left\{\fxa(\xa,\ct/b^\star)\,\fxb(\xb,\ct/b^\star)
\exp\sud(Q,b^\star)+
\left(q \leftrightarrow \bar{q} \right)\right\}\,, \phantom{spingi}
\end{eqnarray}
where
\beq
b^\star=\frac{b}{\sqrt{1+b^2/\blim^2}}\;.
\eeq
The new variable $b^\star$ is approximately equal to $b$ for small $b$,
but never exceeds $\blim$. We should have
$\blim\lesssim 1/\LambdaQCD$, so that
both the Sudakov exponent (\ref{suddef})
and the parton densities do not become undefined because of the Landau
pole.

The region of large $b$ is in any case out of the control
of perturbation theory, since it involves large values of the
strong coupling constant. Therefore, one might want to keep $\blim$
substantially smaller than the value corresponding to the Landau pole,
and parametrize non-perturbative effects by modifying the form factor
in a way that leaves it unaffected at small $b$.
A procedure that is often adopted is to
multiply the integrand of eq.~(\ref{ffdef}) by
\beq
\label{fnp}
F_{NP}=e^{-gb^2},
\eeq
where the parameter $g$ can be adjusted to the data.
Notice that at large $b$, $b^\star$ approaches $\blim$, and therefore
$\exp\sud$ approaches a constant value while
$\exp(ib\qt)$ oscillates. Nevertheless, the integral is convergent
even in the limit $g\,\to\, 0$. In fact, if we add and subtract the
contribution of $\exp\sud$ at $b=\blim$ we obtain
\beqn
\label{ff2}
F(Q,\qt)&=&
\frac{Q^2}{4\pi}\int d^2 b \,
e^{i\vec{q}_{\sssrm T}\cdot\vec{b}}\,e^{-gb^2}
\left[\exp\sud(Q,b^\star)-\exp\sud(Q,\blim)\right]
\nonumber\\
&&+\frac{Q^2}{4g}e^{-\qtq/(4g)}\exp\sud(Q,\blim).
\eeqn
For $g=0$ the integral is now convergent at $b\to +\infty$ for any choice
of $\blim$, and for the remaining term we have
\begin{equation}
  \lim_{g\to 0} \frac{Q^2}{4g}e^{-\qtq/(4g)}\exp\sud(Q,\blim)
        = Q^2 \pi\delta^2(\vec{q}_{\sssrm T})\exp\sud(Q,\blim)\;.
\end{equation}
This fact may be useful to assess the relevance of the non-perturbative
term compared to the effects of resummation.

\end{document}